\documentclass[%
 reprint,%
 amssymb, amsmath,%
 aip,cha,%
]{revtex4-1}

\usepackage{bm}%
\usepackage[colorlinks=true,linkcolor=blue,citecolor=blue,urlcolor=blue]{hyperref}%
\usepackage{graphicx}
\expandafter\ifx\csname package@font\endcsname\relax\else
 \expandafter\expandafter
 \expandafter\usepackage
 \expandafter\expandafter
 \expandafter{\csname package@font\endcsname}%
\fi
\usepackage[outdir=./]{epstopdf}
\newcommand{\bea}{\begin{eqnarray}}
\newcommand{\eea}{\end{eqnarray}}
\newcommand{\be}{\begin{equation}}
\newcommand{\ee}{\end{equation}}

\newcommand{\rmg}{\mathrm{g}}

\newcommand{\rmqp}{\mathrm{qp}}
\newcommand{\rmpp}{\mathrm{p}}
\newcommand{\kBT}{k_\mathrm{B}T}
\newcommand{\rmB}{\mathrm{B}}

\newcommand{\rmT}{\mathrm{T}}

\begin{document}

\title{Correlation-induced refrigeration with superconducting single-electron transistors}%

\author{Rafael S\'anchez}%
\affiliation{Instituto Gregorio Mill\'an Barbany, Universidad Carlos III de Madrid, 28911 Legan\'es, Madrid, Spain}%

\date{\today}%

\begin{abstract}
A model of a superconducting tunnel junction which refrigerates a nearby metallic island without any particle exchange is presented. Heat extraction is mediated by charge fluctuations in the coupling capacitance of the two systems. The interplay of Coulomb interaction and the superconducting gap reduces the power consumption of the refrigerator. The island is predicted to be cooled from lattice temperatures of 200~mK down to close to 50~mK, for realistic parameters. The results emphasize the role of non-equilibrium correlations in bipartite mesoscopic conductors. This mechanism can be applied to create local temperature gradients in tunnel junction arrays or explore the role of interactions in the thermalization of non-equilibrium systems.
\end{abstract}

\maketitle


Quantum coherent processes relevant to quantum information or quantum thermodynamics~\cite{pekola_towards_2015} are usually damaged by temperature. Solid state quantum simulators~\cite{georgescu_quantum_2014} or heat engines~\cite{bjorn_review,benenti_fundamental_2017}
have recently been proposed that operate at very low temperatures. Finding ways to cool nanoscale systems~\cite{giazotto_opportunities_2006,courtois_electronic_2014}, or introduce local temperature gradients well below 100~mK without injecting charge into them is hence a demanding task. This way they are minimally perturbed and not affected by Joule heating.

Thermoelectric refrigeration is traditionally based on the Peltier effect. Driving a current through a conductor with broken electron-hole symmetry converts thermal excitations into transport. Mesoscopic refrigerators based on this effect make use of semiconductor quantum dots~\cite{edwards_quantum_1993,prance_electronic_2009}, superconducting-insulator-normal metal (SIN) junctions~\cite{nahum_electronic_1994,leivo_efficient_1996,pekola_limitations_2004,mastomaki_inas_2017}, or single-electron transistors (SET)~\cite{pekola_refrigerator_2014,feshchenko_experimental_2014}.  Cooling mediated by the coupling to cavity photons has also been proposed in Josephson junctions ~\cite{chen_quantum_2012,karimi_otto_2016,hofer_autonomous_2016} and implemented for the refrigeration of metallic islands~\cite{timofeev_electronic_2009,meschke_single-mode_2006,partanen_quantum_2016,tan_quantum_2017,silveri_theory_2017}.

\begin{figure}[b]
\includegraphics[width=\linewidth]{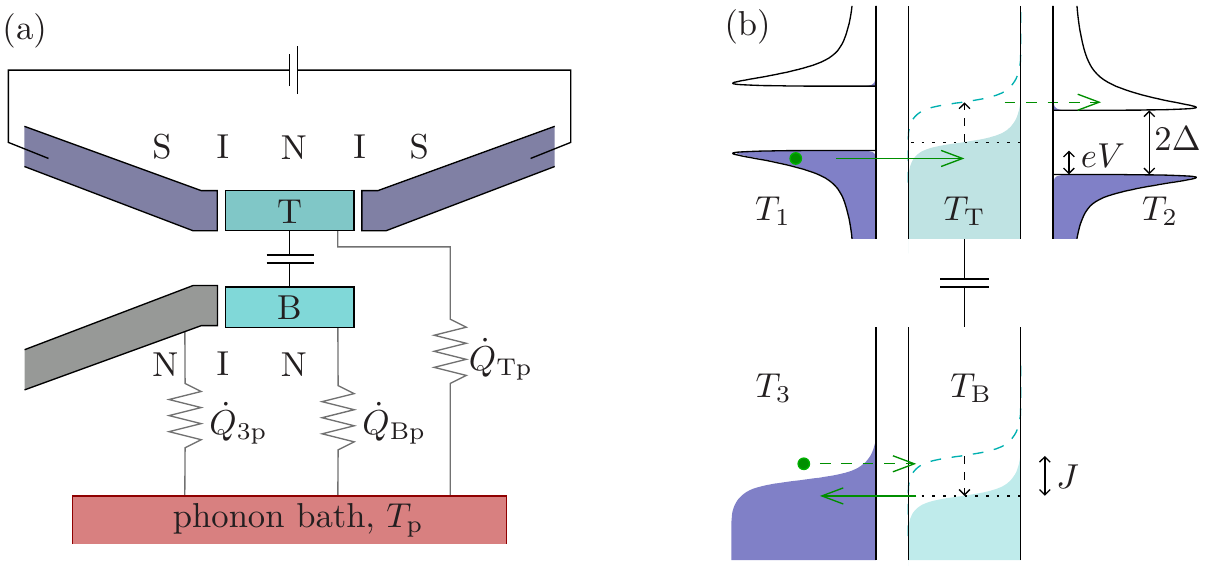}
\caption{\label{fig:scheme}Refrigeration by coupled fluctuations. (a) The electric current through a SINIS SET is coupled to the charge of a SEB. (b) Electrons tunneling into the SET can only overcome the superconducting gap, $\Delta$, by absorbing an energy $J$ from the charging of the SEB. As a result the SEB gets cooled. The temperature of each part of the circuit, $T_i$ is determined by the thermalization with a phonon bath at temperature $T_{\rm p}$.}
\end{figure}

Here an alternative approach is proposed based on the capacitive coupling of a two-terminal SINIS SET to the system to be cooled, see Fig.~\ref{fig:scheme}. We consider cooling of a single-electron box (SEB)~\cite{markusSEB,lafarge_direct_1991}: a Coulomb blockade island which exchanges electrons with a third terminal. The extraction of energy is mediated by the Coulomb repulsion of electrons in different islands, $J$.
The superconducting gap, $\Delta$, acts as an energy filter: At subgap voltages, only transitions that take energy from from the capacitor contribute to transport. A normal metal SET would unavoidably emit Joule heat at all voltages into the island. The manipulation of electrical and thermal flows in capacitively coupled normal metal or semiconductor systems has been recently demonstrated, including effects such as 
quantum dot heat engines~\cite{hotspots,holger,cavities,roche_harvesting_2015,zhang_three-terminal_2015,rob,dare_powerful_2017,walldorf_thermoelectrics_2017}, autonomous Maxwell's demon operations~\cite{strasberg_thermodynamics_2013,koski_onchip_2015}, thermal rectifiers~\cite{ruokola_single_2011,thierschmann_thermal_2015,zhang_three_2017}, memristors~\cite{li_double_2017}, or single-electron current switches~\cite{singh_distribution_2016,transistor,devices}. 

The interplay of the two relevant energy scales, $J$ and $\Delta$, in the cooling mechanism is investigated. Subgap transport in the SET is assisted by the Coulomb interaction, hence increasing the voltage window where cooling takes place. Hot electrons in the SET island are thus pumped over the gap of the superconductor drain due to the charging of the SEB, and replaced by cold electrons from the other contact, see Fig.~\ref{fig:scheme}(b). As a result, both islands get cooled. While the refrigeration of the SET improves for smaller $J$, optimal values can be found for each configuration of the SEB.  

Due to strong Coulomb interactions, the number of electrons in each island is quantized. The electrostatics of the coupled system is described by the model~\cite{koski_onchip_2015}
\be
H=E_\rmT(n-n_\rmg)^2+E_\rmB(N-N_\rmg)^2+J(n-n_\rmg)(N-N_\rmg),
\ee 
where $E_i$ is the charging energy of SINIS SET ($i$=T, for transistor) and the SEB islands ($i$=B, for box), $J$ is the interdot charging energy. The charge of the SET and SEB islands, containing $n$ and $N$ extra electrons, is controlled by $n_\rmg$ and $N_\rmg$, respectively. These can be externally tuned via plunger gates.
We restrict ourselves to a strong Coulomb blockade configuration with up to one extra electron in each island, i.e. $n,N=0,1$. Tunneling in rates through terminal $l$ into island $i$ are quite generally described by: 
\begin{align}
\label{rates}
\gamma_{nN,l}^{+}[\xi]{=}\frac{1}{e^2R_l}\int {dE}{\xi}{\cal N}_{l}(E)f_l(E)[1{-}f_i(E{-}U_{nN}^{l})]
\end{align}
with the associated resistances $R_l$, and the Fermi functions $f_i(E)=\left(1+e^{E/\kBT_i}\right)^{-1}$. For tunneling out rates, $\gamma_{nN,l}^{-}(\xi)$, one needs to replace $f_i(E)\to1-f_i(E)$ in Eq.~\eqref{rates}. We have introduced the parameter $\xi$ which will be different for the different charge or heat processes, see below. The chemical potential of each island is increased by $J$ upon the occupation of the other one:
\begin{align}
\label{U0N}
U_{0N}^l&=2E_\rmT\left(\frac{1}{2}-n_\rmg\right)+J(N-N_\rmg)-\frac{(-1)^l}{2}eV,
\end{align} 
for tunneling between $l$=1,2 and the SET island, and
\begin{align}
\label{Un0}
U_{n0}^3&=2E_\rmB\left(\frac{1}{2}-N_\rmg\right)+J(n-n_\rmg),
\end{align} 
for the SEB.  
In order to take into account the presence of subgap states in the superconducting leads, one considers the Dynes density of states:~\cite{pekola_environment_2010}
\be 
{\cal N}_{1,2}(E)=\left|{\rm Re}\left(\frac{E+i\gamma}{\sqrt{(E+i\gamma)^2-\Delta^2}}\right)\right|,
\ee
with a phenomenological inverse quasiparticle lifetime, $\gamma$. $\Delta$ is the superconducting gap. For the normal metal lead, ${\cal N}_3(E)=1$.


The particle tunneling rates $\gamma_{nN,l}^\pm[\xi{=}1]$ define the charge occupation of the system, $p(n,N)$, via a set of four rate equations~\cite{averin_single_1991,hotspots}. 
We are interested in the stationary regime described by the solution of $\dot p(n,N)=0$. 
With this the different currents are written in terms of the following expressions:
\begin{align}
{\cal I}_l[\xi]&=\sum_N\left[\gamma_{0N,l}^+[\xi]p(0,N)-\gamma_{1N,l}^-[\xi]p(1,N)\right]\\
{\cal I}_3[\xi]&=\sum_n\left[\gamma_{n0,1}^+[\xi]p(n,0)-\gamma_{n1,3}^-[\xi]p(n,1)\right].
\end{align}
With this notation, the charge current through the SET is $I_1=-I_2=e{\cal I}_1[\xi{=}1]$. In the SEB, $I_3=0$. The  heat currents injected from terminals $l$=1,2,3 are then $\dot Q_l={\cal I}_l[\xi{=}E]$.
The heat flowing out of the two islands, T and B, reads:
\begin{align}
\dot Q_\rmT=&-{\cal I}_1[\xi{=}E{-}U_{nN}^1]-{\cal I}_2[\xi{=}E{-}U_{nN}^2]\\
\dot Q_\rmB=&-{\cal I}_3[\xi{=}E{-}U_{nN}^3].
\end{align}

One also needs to assume that the different parts of the system exchange heat, $\dot Q_{i\rmpp}$, with a thermal bath (phonons) at temperature $T_{{\rm p}}$, as sketched in Fig.~\ref{fig:scheme}(a). The final piecewise temperature distribution achieved by the refrigerator is obtained by solving the system of coupled balance equations: 
\be
\label{balance}
\dot Q_{i}(V,T_i)=\nu_i\Sigma(T_{{\rm p}}^5-T_i^5),
\ee
for the normal metal parts, $i{=}$T,B,3. The coupling depends on the volume of the metal parts, taken here as $\nu_i=30\times50\times300$~nm$^3$, and the material-dependent constant, $\Sigma=2\times10^9$~Wm$^{-3}$K$^{-5}$. 
This set of equations is highly nonlinear and need to be solved numerically. The results below show the solutions $\{T_\rmT,T_\rmB,T_3\}$ obtained for some experimentally relevant configurations.
We assume that the superconducting leads are coupled to quasiparticle traps such that they do not overheat~\cite{nguyen_trapping_2013}. In order to proceed, an effective temperature $T_1=T_2=T_\rmqp=300$~mK is considered, independently of the applied voltage $V$.

\begin{figure}
\includegraphics[width=\linewidth]{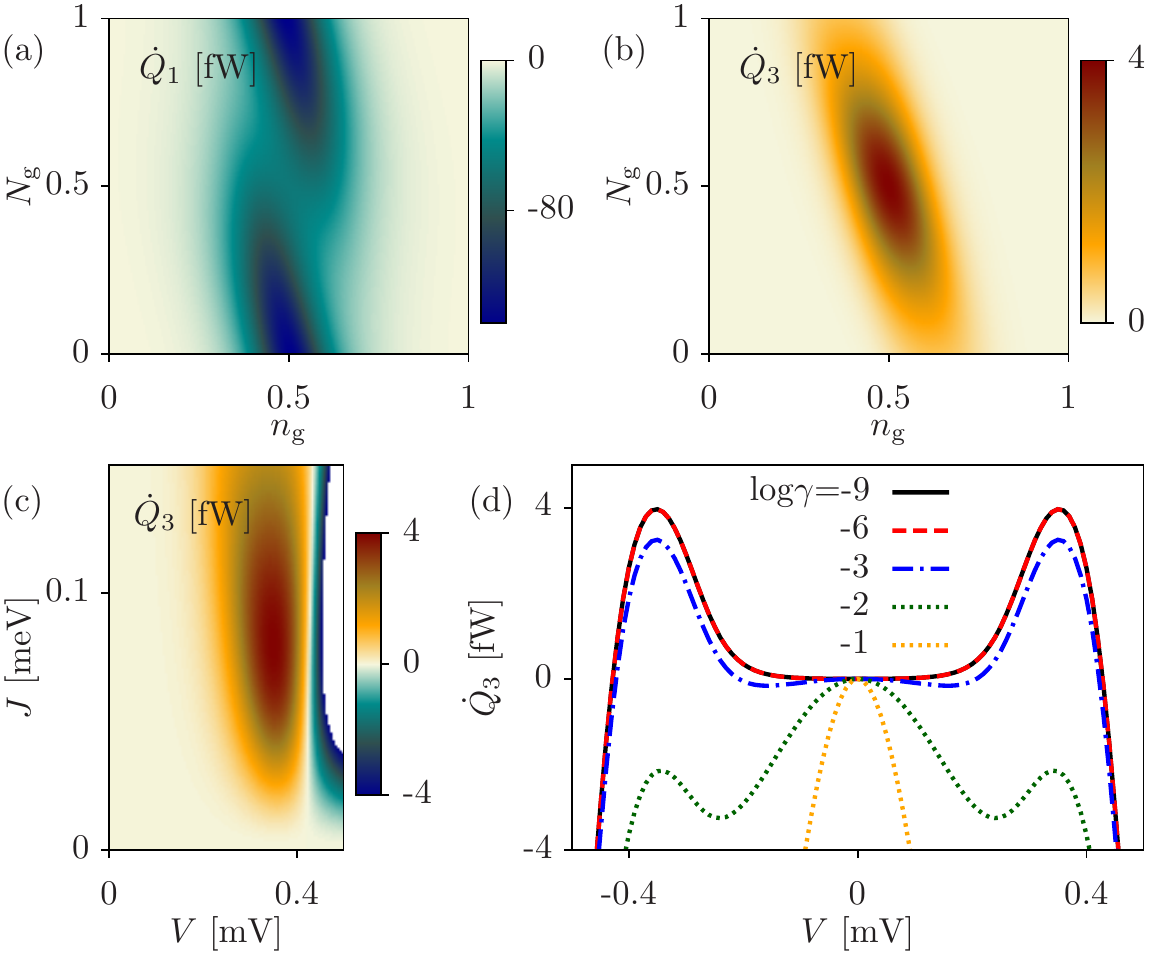}
\caption{\label{fig:heatcurr}Heat currents through (a) terminal 1 and (b) terminal 3 as functions of the gating $n_{\rm g}$ and $N_{\rm g}$. Here $T$=200~mK, $V$=0.35~mV, $J$=0.08~meV, $\Delta$=0.2~meV and $R_i=R_0$=10~k$\Omega$. (c) Heat extracted from the SEB by tuning the applied voltage $V$ and the interaction $J$. (d) Cooling of the SEB is robust up to quasiparticle lifetimes of the order $\gamma>10^{-3}$. }
\end{figure}

The refrigeration is assisted by the correlation of charge fluctuations in both islands, which are maximal around the configuration with $n_\rmg=N_\rmg=1/2$~\cite{correlations,koski_onchip_2015}. Sequences of the type ${(0,0)}\rightarrow{(1,0)}\rightarrow{(1,1)}\rightarrow{(0,1)}\rightarrow{(0,0)}$ involve two tunneling processes in the SET island which occur at different occupations of the SEB, cf. Fig.~\ref{fig:scheme}(b). The tunneling in and out events in both island occur at potentials that differ by the charging energy. An energy $U_{10}^3-U_{00}^3=J$ is  extracted from the SEB~\cite{hotspots}, see Eqs.~\eqref{U0N} and \eqref{Un0}. This is visible in Fig.~\ref{fig:heatcurr}(a) and (b). Far from the working condition, the superconductor works as a usual SINIS refrigerator~\cite{nahum_electronic_1994,leivo_efficient_1996} around $n_\rmg=1/2$. This behaviour is distorted close to $N_\rmg=1/2$, where $\dot Q_3$ is maximal. 

The interplay of the different energy scales, $eV$, $\Delta$, and $J$ is shown in Fig.~\ref{fig:heatcurr}(c). The refrigerator works around the condition $eV\approx2\Delta-J$, where subgap current is enabled in the SINIS via the  absorption of energy from the SEB. However, there is an optimal value for the interaction: if $J\gg\kBT$, the transition $(1,0)\rightarrow(1,1)$ is suppressed and the two systems do no longer exchange energy~\cite{transistor}. The superconductor gaps act as filters that avoid electrons from tunneling through the  SET without exchanging energy with the SEB. This property is limited by subgap states. However the refrigeration is robust up to values of $\gamma$ larger than those typically found in experiments ($\gamma\sim10^{-5}$), cf. Fig.~\ref{fig:heatcurr}(d). 

\begin{figure}
\includegraphics[width=\linewidth]{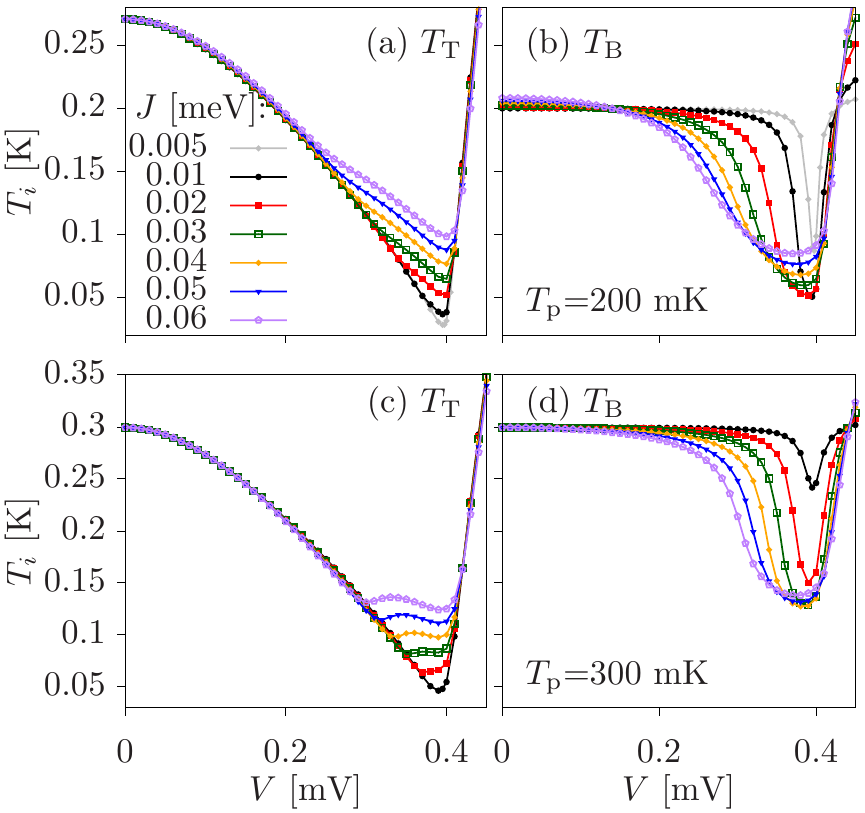}
\caption{\label{fig:TI3}Temperature in the islands as a function of the applied voltage, for different interactions $J$. The superconducting leads are at $T_\rmqp=300$~mK, whereas the phonon bath is at: (a),(b) $T_\rmpp=200$~mK and (c),(d) $T_\rmpp=300$~mK.  The rest of parameters are as in Fig.~\ref{fig:heatcurr}, with $n_\rmg=N_\rmg=1/2$ and $\gamma=10^{-5}$.}.
\end{figure}

The temperature of the SEB is not affected by a small applied voltage, but drops abruptly in the region ${2\Delta-J}<eV<2\Delta$, as expected, see Fig.~\ref{fig:TI3}. Apart from increasing the size of the voltage window at which cooling takes place, the interaction affects the lowest achievable temperature, $T_{\rm B,m}$. As $\dot Q_3$ is proportional to $J$~\cite{hotspots}, for small interaction, the extracted current is small and cooling is not effective. The optimal $J$ is found to depend non-trivially on the bath temperature. As shown in Figs.~\ref{fig:TI3}(b),(d), the minimal temperature is obtained for $J\approx0.01$~meV at $T_\rmpp=200$~mK, and for $J\approx0.03$~meV at $T_\rmpp=300$~mK, for otherwise similar configurations. Remarkably in the former case, the SEB is cooled from $T_\rmB=200$~mK down to $T_{\rm B,m}\approx50$~mK, for realistic parameters. For larger voltages, transport is open in the SINIS for processes that inject energy into the SEB, which is then heated up. 

One might not be interested in finding the lowest possible temperature, but rather in cooling the system down to a target temperature, say $T_\rmB^*$. In this case, the effect of larger $J$ reduces the required power consumption, for achieving $T_\rmB^*$ at lower voltages.

On the other hand, the coupling to the SEB is always detrimental for cooling the SINIS island, see Figs.~\ref{fig:TI3}(a),(c). The onset of transport assisted by fluctuations in the SEB injects energy in the conductor. This is visible as a secondary feature in the dip whose position moves to lower voltages for larger $J$. This feature manifests as a shoulder for $T_\rmpp=200$~mK and as a second dip for $T_\rmpp=300$~mK.

Due to the opposite $J$-dependence in the two subsystems, it is found that $T_{\rm B,m}<T_{\rm T,m}$ for large enough interaction. This would be surprising if the SEB island was a passive system which is cooled down just by being in thermal contact with the refrigerator. In that case, its temperature should always be ${\min}(T_\rmT,T_\rmpp)<T_\rmB<{\max}(T_\rmT,T_\rmpp)$. However, SEB cannot be considered as a separate system. The two islands form a bipartite configuration whose fluctuations are strongly correlated, and maintained out of equilibrium by the voltage bias. This way heat can actively be extracted from the SEB even if the SINIS island to which it is coupled is hotter. 

\begin{figure}
\includegraphics[width=\linewidth]{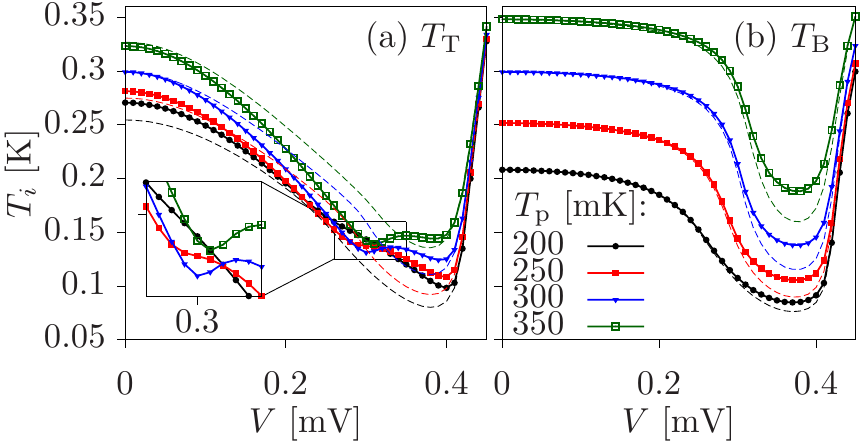}
\caption{\label{fig:Tbprobe}Temperature of the two islands for the same configuration of Fig.~\ref{fig:TI3}(a),(b), with $J=0.06$~meV, and different bath temperatures. The dashed lines show the temperature $T_i^{\rm th}$ measured by temperature probes coupled to the islands (with resistances $R_{\rm th}=100R_0$).}
\end{figure}

To further explore this effect, Fig.~\ref{fig:Tbprobe} shows the two island temperatures, for an increasing temperature of the phonon bath. Recall that the superconductor is assumed to be fixed at $T_\rmqp=300$~mK. Clearly, for large enough temperatures of the thermal bath, the contribution of transport-induced fluctuations is suppressed, and one always finds $T_{\rm T,m}<T_{\rm B,m}$, as expected for systems in thermal equilibrium locally. 

The temperature inversion discussed above might be attributed to the question of what is the temperature of a system which is subject to strong (voltage) fluctuations~\cite{stafford_local_2016}. This is a fundamental question out of the scope of this letter. However, as a first attempt to clarify this issue, a model for temperature probes is introduced explicitly. This is done by assuming fictitious terminals weakly coupled to the two islands whose temperature and voltage, $T_i^{\rm th}$ and $V_i^{\rm th}$, adapt to the conditions that the probes inject no charge and no energy into the system~\cite{markus_probe,stafford_local_2016}. The system of equations, $I_i^{\rm th}=0$ and $\dot Q_i^{\rm th}=0$, for $i$=T,B, is solved using the temperatures obtained by Eq.~\eqref{balance} as input. Note that under these conditions, heat currents are balanced in the islands, i.e. they absorb no net heat. Any difference $T_i-T_i^{\rm th}$ is hence only due to non-equilibrium fluctuations. As shown in Fig.~\ref{fig:Tbprobe}, the measured temperature is indeed different from that obtained from the thermalization with the phonon bath. In all cases, it is obtained $V_\rmT^{\rm th}=V_\rmB^{\rm th}=0$. However, the measured $T_\rmB^{\rm th}$ can still be lower than $T_\rmT^{\rm th}$. One can therefore rule this interpretation out.

Note also that for some intermediate voltages, cooling the SINIS island is improved by having hotter phonons, as shown in the inset of Fig.~\ref{fig:Tbprobe}(a). This is due to the behaviour of the $J$-dependent secondary feature, which develops as a dip at larger temperatures. This feature is smeared in the probe response, such that $T_\rmT^ {\rm th}$ increases monotonically with $T_\rmpp$ for a given $V$. 

\begin{figure}
\includegraphics[width=0.6\linewidth]{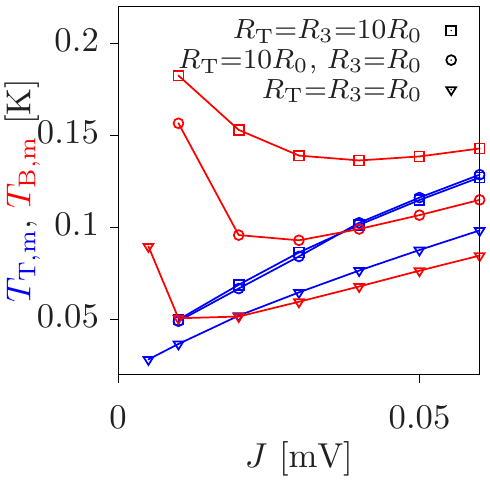}
\caption{\label{fig:mintemp}Minimal achieved temperature in the islands, depending on the interaction $J$ and the tunnel resistances, with $R_1=R_2=R_\rmT$ and $R_0=10$~k$\Omega$. Here, $T_\rmpp=200$~mK.}
\end{figure}

Finally, the effect of the tunneling resistances is considered in Fig.~\ref{fig:mintemp}. The temperature of the SINIS island is almost independent of the resistance of the SEB barrier, $R_3$. Differently, the  minimal temperature achieved in the SEB is strongly affected by all barriers, as it depends on the correlation of tunneling events in the two systems. Increasing the resistance of the SINIS barriers $R_\rmT$ one order of magnitude reduces the temperature gradient in around 1/3, while still obtaining temperatures around $T_\rmB\approx0.1$~mK. 

To conclude, a mechanism for cooling a nanoscale system is proposed which requires no electric exchange. The system and a nanostructured refrigerator are capacitively coupled and form a bipartite system whose fluctuations are strongly cross-correlated. A model based on a coupled SINIS-SEB is considered. Transport at subgap voltages is enabled by charge fluctuations in the SEB involving the transfer of energy via electron-electron interactions. As a consequence, both islands are cooled from lattice temperatures of 200~mK down to around 50~mK. This way, local temperature gradients can be maintained in nanostructured conductors showing peculiar transport phenomena~\cite{detailedbalance}. The effect of the Coulomb interactions is investigated, leading to unexpected configurations where the system is colder than the refrigerator island. These findings suggest the impact of non-equilibrium correlations on the thermodynamic properties of mesoscopic configurations which are under nowadays experimental reach~\cite{feshchenko_experimental_2014,koski_onchip_2015,singh_distribution_2016}.  

I am grateful to Jukka P. Pekola for encouraging discussions and his valuable help, and thank the Aalto University for its hospitality during a visit when this work was conceived. This work was supported by the Spanish Ministerio de Econom\'ia, Industria y Competitividad (MINECO) via grant FIS2015-74472-JIN (AEI/FEDER/UE). I acknowledge the MINECO MAT2016-82015-REDT network.

\bibliography{papersinis}

\end{document}